\documentclass[11pt,twoside,onecolumn]{article}
\usepackage{amsmath}
\usepackage{graphicx,epsf}
\usepackage[dvips]{epsfig}
\newcommand{\be}{\begin{eqnarray}}
\newcommand{\ee}{\end{eqnarray}}

\title{\bf The Equation of State of an Interacting Tachyon}
\author{G.L. Alberghi\thanks{e-mail: alberghi@bo.infn.it},\\
{\em Physics and Astronomy Department, Bologna University} \\
{\em and} \\
{\em Istituto Nazionale di Fisica Nucleare, Sezione di Bologna,
Italy} \\
\\
A. Tronconi \thanks{e-mail: tronconi@bo.infn.it}  \\
{\em Physics Department, Bologna University} \\
{\em and} \\
{\em Istituto Nazionale di Fisica Nucleare,
Sezione di Bologna, Italy}}
\begin{document}
%
%
\maketitle
\begin{abstract}

We examine the cosmological solutions of a tachyon field non
minimally coupled to gravity through an effective Born-Infeld interacting Lagrangian with
a power law potential, in order to investigate its equation of state as related to tracking properties.
We find exact solutions in the case of a tachyon dominated universe and when
the dominant component of the stress energy tensor is determined by some other
perfect fluid.

\end{abstract}
%
\pagestyle{plain}
\raggedbottom
\setcounter{page}{1}
\section{Introduction}
\setcounter{equation}{0}
%
%



%
%
%
%

Experimental observations performed in the last few years in the
context of Cosmic Microwave Background \cite{Bennett} and of the
type Ia Supernovae \cite{SN} strongly suggest that the Universe
has been going through a phase of accelerated expansion for the
last 5-10 Gy. The main candidates usually introduced in order to
explain this behavior are a cosmological constant and a scalar
field which begins to dominate at the latest cosmological phase.
One of the possibilities taken into account in the literature is
that scalar field is a tachyonic one. This kind of field appears
in the context of string theory \cite{Sen} and has been
extensively applied to Cosmology \cite{Gibbons}. Usually the
tachyon is taken without explicit coupling to gravity (in analogy
with a minimally coupled ordinary scalar field). We will extend
the discussion of the properties of the system when an explicit
non-minimal coupling is introduced. In particular we will be
interested in solutions of the field equations in the case when
the tachyon stress energy density is the dominant part of the
content of energy of the Universe and, on the
opposite, when the total stress energy density is dominated (and
coherently the scale factor evolution) by some other fluid. In
particular we will discuss the equation of state of the tachyon
field, as it is at the base of the determination of the tracking
feature (\cite{Scherrer, Aguirregabiria}). We will start with a
Born-Infeld Lagrangian assuming a power-law tachyon potential.
This will be done in analogy to the analysis of \cite{Abramo} in
order to determine the qualitative and quantitative effects of the
gravitational interaction term. In fact one of the many
interesting results obtained in \cite{Abramo} is the presence of
two attracting solution with a dust and a cosmological constant
like behavior. We will look at the solutions of the coupled
equations of motion for the tachyon and for the scale factor of a
FRW metric in order to explicitly determine the behavior of the
field and thus its equation of state to see if the tracking
behavior is still present in the interacting case. In Section 2 we
introduce the basic framework for the tachyon effective action. In
Section 3 we consider the case when the tachyon energy density is
the dominant one in the Friedmann equations and in Section 4 the
case when a different fluid drives the expansion of the scale
factor. In Section 5 we discuss the results.

\section{The Basic Framework}

The action describing the non interacting tachyon field is usually
(see i.e. \cite{Abramo}) taken in a Born-Infeld form: \be
\label{L_BI} {\cal{L}}_{BI} = - \sqrt{-g} \, V({\varphi}) \,
\sqrt{1 - \frac{\partial^\mu \varphi
\partial_\mu \varphi}{M^4}} \; .
\ee This scalar field was proposed in connection with string
theory, as it seems to represent a low-energy effective theory of
D-branes and open strings, and has been conjectured to play a role
in cosmology \cite{Sen,Padma}. The energy and pressure densities
of the tachyon, in the homogeneous case, are:
\begin{eqnarray}
\label{rho_T}
\rho_T &=&  \frac{V(\varphi)}{\sqrt{1 - \frac{\dot \varphi^2}{M^4}}} \; ,\\
\label{p_T}
p_T &=& - V(\varphi) \sqrt{1 - \frac{\dot \varphi^2}{M^4}} \; .
\end{eqnarray}
The tachyon fluid is also characterized by the ratio between pressure and
energy (the equation of state) $w_t$ and sound speed $c_T^2$:
\be
w_T = - 1 + \frac{\dot \varphi^2}{M^4} \quad , \quad c_T^2 = - w_T  \; .
\ee
Since the equation of state is necessarily nonpositive because of the
square root in the action (\ref{L_BI}), the theory is stable
--- energy and pressure are real, and inhomogeneous perturbations
have a positive sound speed.
Moreover, because $w_T \le 0$, the tachyon is a natural candidate
for dark energy and inflation. Another interesting property is
that the equation of state and sound speed of tachyons are equal,
but with opposite signs, irrespective of the form for the
potential. Canonical scalar fields, on the other hand, obey the
Klein-Gordon equations, hence their fluctuations travel with sound
speed equal to unity (in units where $c=1$). Therefore, tachyon
fluctuations are fundamentally different from the fluctuations of
a canonical scalar field, irrespective of the shape of the
potential.

\par
One might expect drastic changes by taking a non-minimal coupling
of $\phi$, as encoded in the Lagrangian \be
 L_{\phi} =  \sqrt{- g} \Big[  - V (\phi) \sqrt{ 1 -
  g^{\mu\nu} \partial_{\mu} \phi \partial_{\nu} \phi  + \xi R \phi^2}\Big],
\label{lagrangian} \ee where $R$ is the Ricci scalar, $\xi $ is
the non-minimal coupling constant and $V(\phi)$ is the potential,
so that the total Lagrangian may be written as
\be
 L_{tot} =  \sqrt{- g} \Big[  - V (\phi) \sqrt{ 1 -
  g^{\mu\nu} \partial_{\mu} \phi \partial_{\nu} \phi  + \xi R \phi^2}  
  + \frac{M^2}{2} R\Big],
\ee
where $M$ is the Planck mass.


In the case of a tachyon living in a Universe where some kind of
matter, effectively described as a perfect fluid, is present,
Einstein's field equations are given as \be
 R_{\mu\nu} - \frac{1}{2}g_{\mu\nu} R = 8 \pi G [ T_{\mu\nu (\phi)} + T_{\mu\nu (m)}]
\label{einstein}
\ee
with energy-momentum tensor components of tachyon and matter as
\be
 T_{\mu\nu (\phi)} = (\rho_{\phi} + p_{\phi})u_{\mu} u_{\nu} - p_{(\phi)}
  g_{\mu\nu}
\label{stress}
\ee
and
\be
 T_{\mu\nu (m)} = (\rho_{m} + p_{m})u_{\mu} u_{\nu} - p_{m}
  g_{\mu\nu}
\ee respectively, where $u^{\mu} = (1,0,0,0)$. The diagonal
components are defined as \be T^{\mu}_{\mu (\phi)} =
(\rho_{\phi}, - p_{\phi}, - p_{\phi}, - p_{\phi}) \ee and
can be obtained from the lagrangian (\ref{lagrangian}) as
 \be
 T_{\mu\nu (\phi)} &=& - V (\phi) [ 1 -
   \bigtriangledown^{\rho} \phi \bigtriangledown_{\rho} \phi  + \xi R
   \phi^2]^{-1/2} \times \Big[ - \bigtriangledown_{\mu} \phi
   \bigtriangledown_{\nu} \phi + \xi R_{\mu\nu} \phi^2 \\ &&  + \xi (
   \bigtriangledown_{\mu} \bigtriangledown_{\nu} - g_{\mu\nu} {\bigtriangledown}^2 ) \phi^2
   - g_{\mu\nu} ( 1 -
   \bigtriangledown^{\rho} \phi \bigtriangledown_{\rho} \phi  + \xi R
   \phi^2) \Big].
\ee
Here $\bigtriangledown_{\mu}$ stands for the covariant derivative and $R_{\mu\nu}$
are Ricci tensor components.
The field equations for $\phi$ are obtained as
\be
{\bigtriangledown}^2 \phi + \frac{2(\bigtriangledown^{\mu} \phi) (\bigtriangledown_{\rho}
  \phi) (\bigtriangledown^{\rho} \bigtriangledown_{\mu} \phi)
- 2 \xi R \phi
\bigtriangledown^{\rho} \phi \bigtriangledown_{\rho} \phi -
\xi \phi^2 g^{\mu\nu}\bigtriangledown_{\mu} R
\bigtriangledown_{\nu} \phi}{2 (1 -
   \bigtriangledown^{\rho} \phi \bigtriangledown_{\rho} \phi  + \xi R
   \phi^2)} \\
 + \xi R \phi + \frac{V^{\prime}}{V} ( 1 + \xi R \phi^2 ) = 0,
\ee
from the lagrangian (\ref{lagrangian}).
Here $V^{\prime} (\phi) = \frac{d}{d \phi} V(\phi)$ and
\be
 \bigtriangledown^2 \phi  = \bigtriangledown^{\rho} \bigtriangledown_{\rho} \phi=
\frac{1}{\sqrt{-g}}\frac{\partial}{\partial x^{\mu}} ( \sqrt{-g}
g^{\mu\nu} \frac{\partial}{\partial x^{\nu}})\phi. \ee 
According to
cosmological observations \cite{Bennett,SN}, we currently live in
a spatially flat and speeding - up universe, such that ${\ddot a}/
a > 0$ for the scale factor $a(t)$, of a
Friedmann-Robertson-Walker metric

\be
 ds^2 = dt^2 - a^2(t) [ dx^2 + dy^2 + dz^2 ].
\label{metric} \ee In the following we will deal only with the
homogeneous mode of the tachyon field and thus assume
 \be
 \phi (x, t) = \phi (t) .
\ee The Friedmann and the acceleration equations are as usually
written in the form
\begin{equation}
\left(\frac{\dot a}{a}\right)^2 = \frac{8\pi G}{3} \rho \quad
\quad  ;  \left(\frac{\ddot a}{a}\right) =- \frac{4\pi G}{3}
(\rho+3p)
\end{equation}
and the equation of motion for the tachyon is

\be
{\ddot \phi} + 3 H {\dot \phi} + \frac{2 {\ddot \phi}{\dot
\phi}^2 - 2 \xi R
  \phi {\dot \phi}^2 - \xi \phi^2 {\dot R}{\dot \phi}}{2 ( 1 - {\dot \phi}^2 +
  \xi R \phi^2 ) } +  \xi R \phi + \frac{V^{\prime}}{V} ( 1 + \xi R \phi^2 ) =
  0 ,
\label{fieldeq} \ee Of course, only two of these three equations
are independent when the universe is driven by  a single source.
In the following we will make the following ansatz for the tachyon
field
\be
  \phi (t) = A t
\label{phiansatz}
\ee
where $A$ is a dimensionless constant. This
allows us to convert differential equations into algebraic ones,
which is the only way we were able to solve the field equations
for a generic value of the gravitational coupling. The potential
is assumed to be in a power-law form 
\be
   V(\phi) = m^{2-\gamma} \phi^{-2- \gamma}
\label{powerlaw}
\ee
where $m$ has the dimension of a mass, which is the one used, in addition to the exponential one, in
all the phenomenological models.

\section{Tachyon Driven Universe}

In a Friedmann Universe, with the assumptions discussed in the
previous section, the energy density and pressure of the tachyon
field are given by
\begin{align}
\rho_\phi =&-\frac{V(\phi)}{\left[6\xi \phi^2\left(\dot H+2H^2\right)-1\right]\sqrt{1-6\xi \phi^2\left(\dot H+2H^2\right)-\dot \phi^2}}\left\{1+3\xi\phi\left[36\xi \phi^3H^4\right.\right.\nonumber\\
&\left.\left. +3\phi\dot H\left(2\xi \phi^2 \dot H-1\right)+\phi \left(18 \xi \phi^2 \dot H +3 \dot \phi^2-7\right)-H\left(2\dot \phi+3\xi \phi^3 \ddot H\right)H^2\right]\right\}
\end{align}
and
\begin{align} 
p_\phi=&-\frac{V(\phi)}{\left[6\xi \phi^2\left(\dot H+2H^2\right)-1\right]\sqrt{1-6\xi \phi^2\left(\dot H+2H^2\right)-\dot \phi^2}}\left\{15552\xi^4 H^8\phi^8\right.\nonumber\\
&\left.-5184(\gamma+1)\xi^4 H^7\phi^7\dot \phi+432 \xi^3H^6\phi^6\left[-13+8(\gamma+1)\xi+66\xi\phi^2\dot H\right.\right.\nonumber\\
&\left.\left.+(6+4(2\gamma^2+\gamma-1)\xi)\dot\phi^2\right]+\left(\dot \phi^2-1\right)\left[-1+2(\gamma+2)\xi+\left(1+(2\gamma^2+7\gamma+8)\xi\right)\dot \phi^2\right]\right.\nonumber\\
&\left.-324(\gamma+1)\xi^4\phi^7\dot \phi \dot H^2\ddot H-54\xi^3\phi^5\dot \phi \dot H\ddot H\left(\gamma\dot \phi^2-2\gamma-5\right)+3\xi^2\phi^3\dot \phi\ddot H \left[(8+3\gamma)\dot \phi^2\right.\right.\nonumber\\
&\left.\left.-3(\gamma+4)\right]+144\xi^3 \phi^5 H^5\left[\left(14+9\gamma-90(\gamma+1)\xi\phi^2\dot H\right)\dot \phi+(6\gamma-3)\dot \phi^3-18\xi\phi^3\ddot H\right]\right.\nonumber\\
&\left.+12\xi^2\phi^3 H^3\left[\dot \phi \left(-19-9\gamma-756(\gamma+1)\xi^2\phi^4\dot H^2-(12\gamma+13)\dot \phi^2+9\gamma\dot \phi^4+12\xi\phi^2\dot H\right.\right.\right.\nonumber\\
&\left.\left.\left.\times\left(29+15\gamma+3(\gamma-1)\dot \phi^2\right)\right)-6\xi\phi^3\left(7\dot\phi^2-6\right)\ddot H\right]+\xi\phi H\left[-\left(-1+6\xi\phi^2\dot H\right)\left(8+3\gamma\right.\right.\right.\nonumber\\
&\left.\left.\left.+6\xi\phi^2\dot H \left(-35-12\gamma+54(\gamma+1)\xi\phi^2\dot H\right)\right)\dot \phi-2\left(-8-3\gamma+9\xi\phi^2 \dot H\left(-1+2\gamma+6\xi\phi^2\dot H\right)\right)\dot \phi^3\right.\right.\nonumber\\
&\left.\left.+3\left(-8-3\gamma+18\gamma\xi\dot H\right)\dot \phi^5+6\xi\phi^3\left(-3+7\dot \phi^2+6\left(3\xi\phi^2\dot H-\dot \phi^2\right)\left(6\xi \phi^2\dot H+\dot \phi^2\right)\right)\ddot H\right]\right.\nonumber\\
&\left.+54\xi^4\phi^8\dot H\left(12\dot H^3+3\ddot H^2-2\dot H H^{(3)}\right)+\xi\phi^2\left[\dot H\left(-23+60\xi+36\gamma\xi+\left(35+6\right.\right.\right.\right.\nonumber\\
&\left.\left.\left.\left.\times(6\gamma^2+11\gamma+1)\xi\right)\dot \phi^2-12\left(1+2\gamma(\gamma+2)\xi\right)\dot \phi^4\right)-\dot \phi \phi^{(3)}\right]+3\xi^2\phi^4\left[2\dot H^2\left(31\right.\right.\right.\nonumber\\
&\left.\left.\left.-12(4+3\gamma)\xi-2\left(17+3(6\gamma^2+7\gamma-4)\xi\right)\dot \phi^2+6\left(1+\gamma(1+2\gamma)\xi\dot \phi^4\right)-H^{(3)}+4\dot\phi\phi^{(3)}\dot H\right)\right.\right.\nonumber\\
&\left.\left.+9\xi^3\phi^6\left(4\dot H^3\left(-17+12(\gamma+1)\xi+(11+6(-1+\gamma+2\gamma^2)\xi)\dot\phi^2\right)+\left(2\dot\phi^2-3\right)\ddot H^2\right.\right.\right.\nonumber\\
&\left.\left.\left.+4\dot H H^{(3)}-4\dot \phi \phi^{(3)}\dot H^2\right)+36\xi^2 \phi^4H^4\left(21-8(3\gamma+4)\xi-4\left(5+(6\gamma^2+7\gamma-4)\xi\right)\dot\phi^2\right.\right.\right.\nonumber\\
&\left.\left.\left.\left(9+4\gamma(2\gamma+1)\xi\right)\dot\phi^4-36(\gamma+1)\xi^2\phi^3\dot\phi\ddot H+12\xi^2\phi^4\left(57\dot H^2-H^{(3)}\right)+4\xi\phi^2\right.\right.\right.\nonumber\\
&\left.\left.\left.\times\left(-56+36(\gamma+1)\xi+3(5+6(2\gamma^2+\gamma-1)\xi)\dot\phi^2\right)-\dot\phi\phi^{(3)}\right)\right]+3\xi\phi^2H^2 \left[-15\right.\right.\nonumber\\
&\left.\left.+8(3\gamma+5)\xi+\left(22+4(6\gamma^2+11\gamma+1)\xi\right)\dot\phi^2-\left(13+16\gamma(\gamma+2)\xi\right)\dot\phi^4+6\dot\phi^6\right.\right.\nonumber\\
&\left.\left.-432(\gamma+1)\xi^3\phi^5\dot \phi\dot H\ddot H-36\xi^2\phi^3\dot\phi(\gamma\dot\phi^2-2\gamma-5)\ddot H+36\xi^3\phi^6\left(78\dot H^3+3\ddot H^2-4\dot H H^{(3)}\right)\right.\right.\nonumber\\
&\left.\left.+2\xi\phi^2\left(\dot H\left(125-48(3\gamma+4)\xi-4(25+6(6\gamma^2+7\gamma-4)\xi)\dot\phi^2+3(5+8\gamma(2\gamma+1)\xi)\dot\phi^4\right)\right.\right.\right.\nonumber\\
&\left.\left.\left.+4\dot \phi\phi^{(3)}\right)+12\xi^2\phi^4\left(\dot H^2(-119+72(\gamma+1)\xi+6(7+6(2\gamma^2+\gamma-1)\xi)\dot\phi^2)\right.\right.\right.\nonumber\\
&\left.\left.\left.+2 H^{(3)}-4\dot\phi\phi^{(3)}\dot H\right)
\right]
\right\}
\end{align}
where $f^{(3)}(t)\equiv d^3 f(t)/dt^3$.
It is convenient to define a time dependent parameter $w(t)$ by
the relation $w(t) \equiv p_\phi (t) / \rho_\phi (t) $. The
equation of motion for the scalar field, written in the form
$d(\rho a^3) = - w \rho d(a^3)$ can be integrated to give
\be
 (\dot
\rho_\phi/\rho_\phi)= - 3 H (1+w)
\ee
 The Friedmann equation, on
the other hand, gives $\rho_\phi \propto H^2$ so that $(\dot
\rho_\phi/\rho_\phi) = 2 (\dot H/H)$. Combining the two relations
we get
\be
w(t) =-1 - \frac{2}{3} \frac{\dot H}{H^2}
\label{qopw} \ee and thus determine $w(t)$ when the time dependence of the Hubble parameter is known and $\rho_\phi$ is the dominant energy in the Universe. (Note that we have not
used the specific form of the source so far, so this equation
will be satisfied by any source in a FRW model.)

If such a dependence in unknown, from the Einstein equations one can easily write
\be
\label{stateq1} \frac{R^1_1 - \frac{1}{2}R}{R^0_0 - \frac{1}{2}R}
\simeq \frac{- p_{\phi}}{\rho_{\phi}}
\ee
and consider dominance of
tachyon energy over matter.

In this case the Friedmann equation may be solved with our ansatz
for the field $ \phi (t) = A t$ and for the scalar curvature $H(t)=q/t$ only in the case of a potential
with $ \gamma =0 $ and this is that case we will discuss later on.
The equation of state for the tachyon is determined by the
equation
\be
\label{dominantfluid}
  w_\phi \simeq { 2 \dot H + 3 H^2 \over 3 H^2} = { 3q -2 \over 3q}
\ee as derived from the Einstein equation (\ref{einstein}) whenever
one can solve the system of coupled equations
\begin{equation}\label{eq1}
w_\phi = { p_\phi  \over \rho_\phi}
\end{equation}
\begin{equation}\label{eq2}
3A^2q\left[1+2\xi\left(2q-1\right)\left(\gamma+1\right)\right]-\gamma-2=0\end{equation}
for $q$ and $A$.
When $\gamma=0$, in particular, $ \rho_ \phi(\phi) $ and $p_\phi
(\phi) $ in eq. (\ref{eq1}) are given by
\begin{align}
\rho_\phi&=\frac{1-9A^2q\xi(q-1)}{A^4t^4\sqrt{1-A^2\left[1+6q\xi\left(2q-1\right)\right]}},\\
p_\phi&=\frac{-1+A^2\left[1+3\xi\left(3q^2-3q+2\right)\right]}{A^4t^4\sqrt{1-A^2\left[1+6q\xi\left(2q-1\right)\right]}}.
\end{align}

We solved the equations
numerically and the results are quite interesting.
The first aspect to note is that in the limit $ \xi \rightarrow 0 $ 
one recovers the results obtained in the non-interacting tachyon 
(see \cite{Abramo}) as expected.
In fact for every value of the ratio  $ M $ / $ m $,
by taking the limit 
$ \xi \rightarrow 0$ one obtains for $ w_ \phi $ the couple of values $ (-1,0) $.
One can thus infer that the attractor structure of the non interacting
case, where a dust and a cosmological constant-like values are
obtained, is reproduced.
One should then ask what is the effect of the interaction term.
In the case $ M $~/~$m > 1 $ 
(see Fig. 1) 
one can see that there is a maximum value for the coupling constant $ \xi $
in oder to obtain real values for the equation of state.
As one approaches this limiting value for $ \xi $ the cosmological constant and dust
behavior are modified and approach a value of $ w_{\phi} \simeq 0.25 $.

Even in the case $ M $~/~$m < 1 $ 
(see Fig. 2) 
one can see that there is a maximum value for the coupling constant $ \xi $
in oder to obtain real values for the equation of state.
As one approaches this limiting value for $ \xi $ 
both the solutions for the equation of state parameter approach the
cosmological constant case.
One can thus deduce that in this case even as one takes
a very small but fixed value for $ \xi $ there are values of the ratio
$ m $ / $m $ for which the structure of the non intercating case is not
recovered and the only possible tachyon behavior is that of a cosmological 
constant.
One can thus conclude that the introduction of an interaction term
completely modifies the conclusion one might draw in the
non-interacting case even for a very weakly interacting theory.
\par
One would need to introduce a formalism similar to that used in
\cite{Aguirregabiria}, where a phase plane analysis of the flow
defined by the equations of motion was performed. This is beyond
the scope of this preliminary analysis as the coupled interacting
equations cannot be solved analytically as in
\cite{Aguirregabiria}.

\begin{figure}[t]
\label{Fig1}
\centering
\epsfxsize=2.7in
\epsfbox{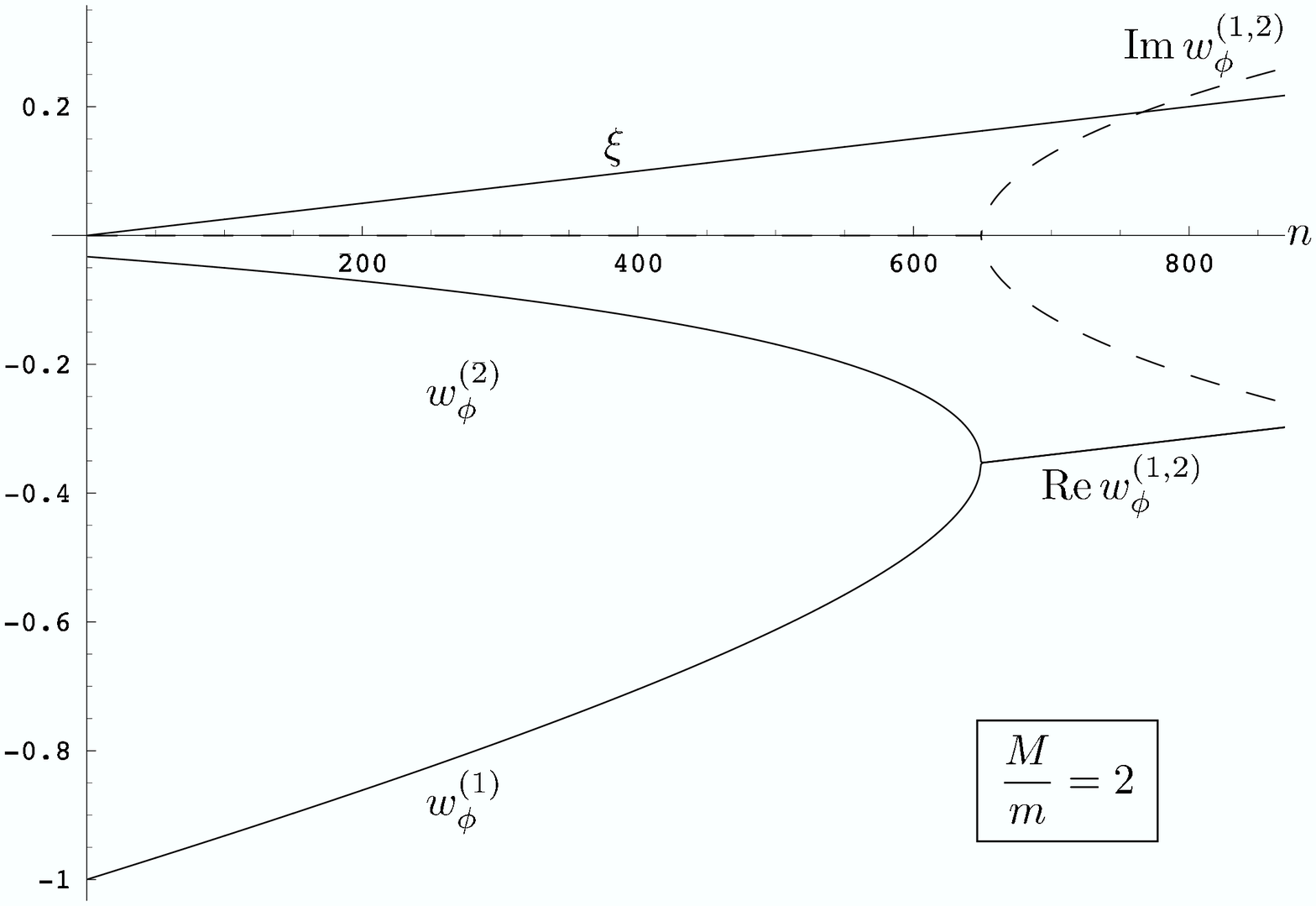}
\epsfxsize=3.2in
\epsfbox{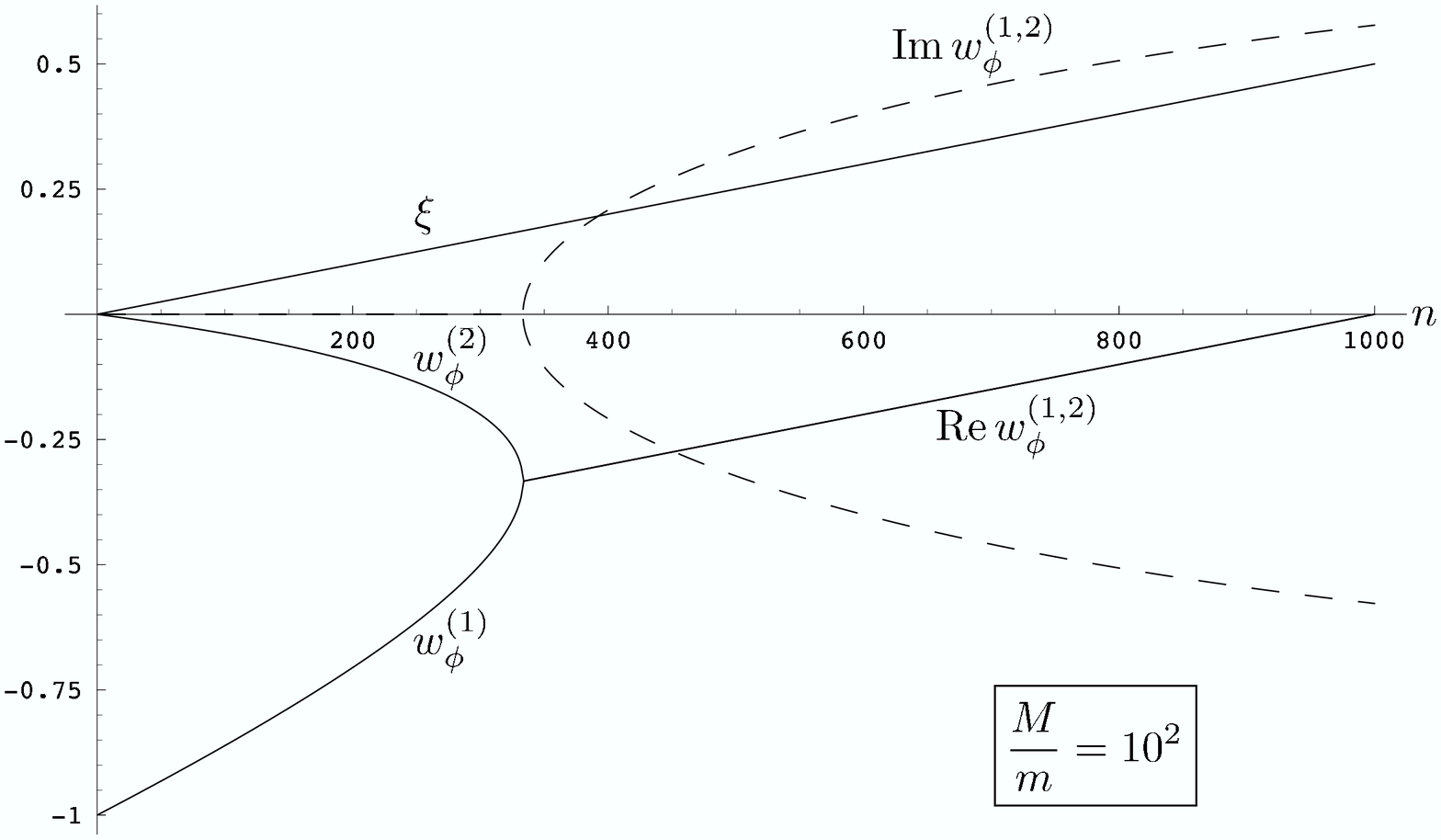}
\caption{\it The two plots above show $w_\phi$ obtained solving eqs. (\ref{eq1}-\ref{eq2}) for different values of $\xi$ when $M/m=2$ (l.h.s) and $M/m=10^2$ (r.h.s). The straight line lying above the horizontal axis in both the plots represents $\xi$ as a function of the discrete index $n$ which labels the solutions obtained. The solid line which bifurcates is the real part of $w_{\phi}$ and the dashed line is the imaginary part of the solutions. Physical solutions are admissible only in the ${\rm Im}w_\phi=0$ region.}
\end{figure}

\section{Tachyon in a Fluid-Driven Universe}

In the case where the Universe expansion is driven by a fluid with equation of state
\be
\label{fluidstate}
   p_f = w_f \rho_f
\ee
for a power-law evolution of the scale factor
\be
   a(t) = a_0 \left({t\over t_o}\right)^q
\ee
one can show that the Hubble parameter is given by
\be
   H(t) = { 2 \over 3(1+w_f) t} = {q \over t}.
\label{hubble1} \ee The equation of motion for the tachion field
in a potential of the form $ V(\phi) = m^{4-\alpha} \phi ^{- \alpha} $
(with $\alpha = 2 + \gamma $ of eq. (\ref{powerlaw}) reduces to
\be
   12 (\alpha -1 ) \xi A^2 q^2 - 3 ( 1+ 2 (\alpha -1 )\xi) A^2 q + \alpha = 0
\ee
where $ q = 2/3(1+w_f) $ as determined by (\ref{hubble1}).
Now one has
\be
  A^2 = - \left. {\alpha \over 12 (\alpha-1) \xi q^2 - 3 [ 1 + 2 (\alpha-1)\xi]q} \; \;
          \right| _ { q= 2/(3(1+w_f)) }
\ee 
By substituting the expression for $A$ in the expression of
the energy density $ \rho _ \phi $ and pressure $ p _ \phi $ one
obtains the equation of state for the tachyon. 
The result of the above procedure is very interesting: 
the equation of state for the tachyon is independent of $\xi$
 \be
   w_ \phi = -1 + \alpha {1 + w_f  \over 2}.
\ee 
The introduced coupling to gravity does not show any effect on
the equation of state for a non-dominant tachyon.

\begin{figure}[t]
\label{Fig2}
\centering
\epsfxsize=3in
\epsfbox{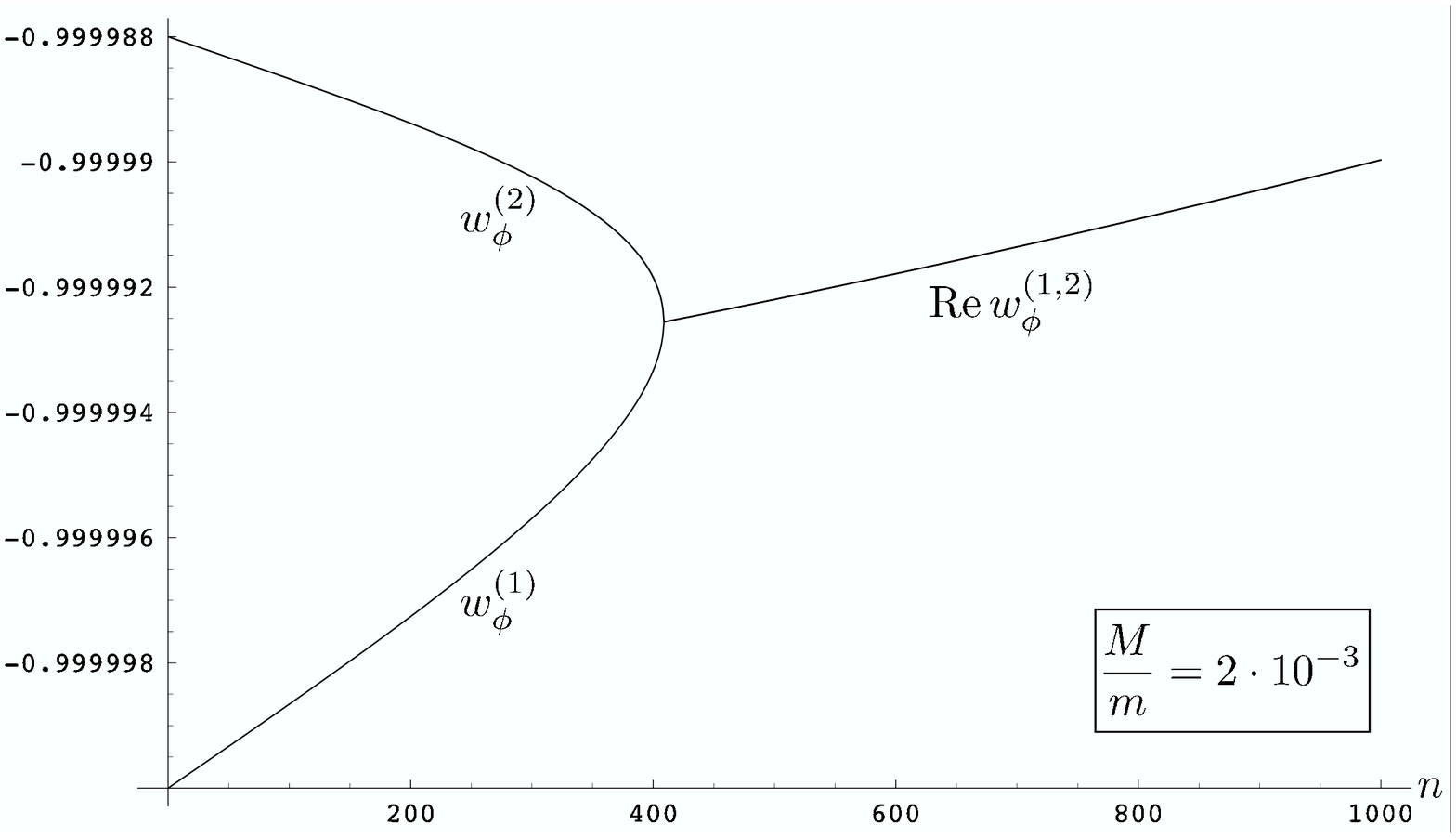}
\epsfxsize=3in
\epsfbox{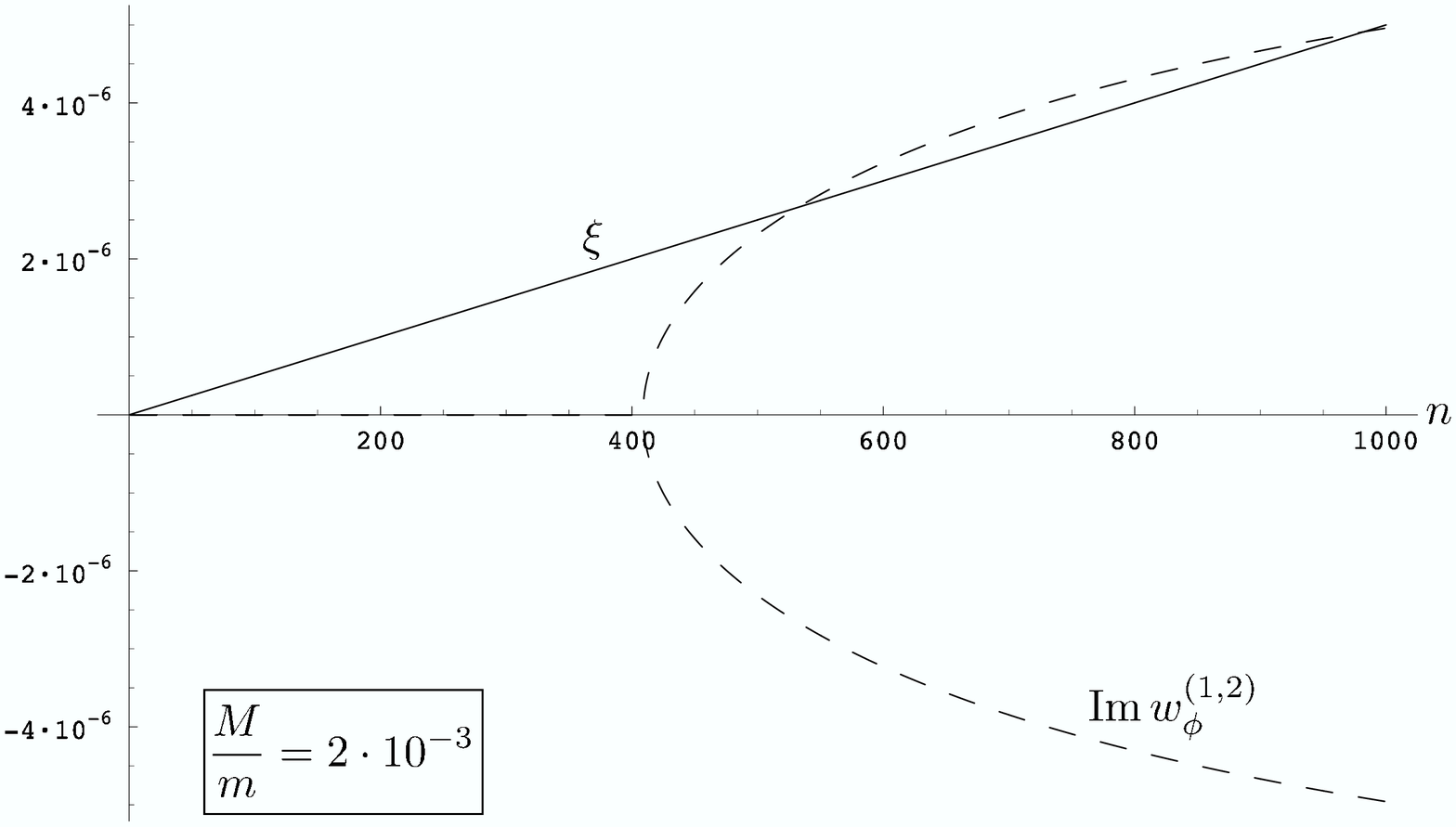}
\caption{\it The two plots above show the real part (l.h.s.) and the imaginary part (r.h.s.)of $w_\phi$ when $M/m=2\cdot 10^{-3}$. The two graphs were obtained solving eqs. (\ref{eq1}-\ref{eq2}) for different values of $\xi$ and the straight line lying above the horizontal axis in the plot on the right represents $\xi$ as a function of the discrete index $n$ which labels the solutions obtained. The solid line which bifurcates in the l.h.s. plot is the real part of $w_{\phi}$ and the dashed line in the r.h.s. plot is the imaginary part of the solutions. Physical solutions are admissible only in the ${\rm Im}w_\phi=0$ region.}
\end{figure}

\section{Conclusion}

We have examined the equation of state for a tachyon field coupled
to gravity through the Born-Infeld Lagrangian (\ref{lagrangian}) in
order to determine the equation of state for a tachyon coupled to
gravity. This generalizes the case described by \cite{Abramo}. 
When the homogeneous tachyon field is the dominant source in the
Friedmann equation, even for small values of the coupling ($\xi$), the
system may behave very differently from the uncoupled case. 
The dust attractor, present in the uncoupled case, may disappear in
favor of a cosmological constant type of behavior. 
On the contrary, when some other fluid is driving the expansion of the
scale factor, the gravitational coupling shows no effect on the
tachyon equation of state.
We feel that these results are quite
intriguing and would need a deeper, thought not simple, numerical 
phase space analysis in order to be fully understood.

\end{document}